# Evaluating Generalizability of Deep Learning Models Using Indian-COVID-19 CT Dataset

Suba S[1[0000-0001-6251-5083]], Nita Parekh[1[0000-0002-8737-0260]], Ramesh Loganathan[1], Vikram Pudi[1[0000-0003-3992-0624]]and Chinnababu Sunkavalli[2]

[1] International Institute of Information Technology, Gachibowli, Hyderabad, 500032, India
[2] Grace Cancer Foundation, Hyderabad, India
suba.s@research.iiit.ac.in,nita@iiit.ac.in,
ramesh.loganathan@iiit.ac.in,
vikram@iiit.ac.in,chinna@gracecancerfoundation.org

**Abstract.** Computer tomography (CT) have been routinely used for the diagnosis of lung diseases and recently, during the pandemic, for detecting the infectivity and severity of COVID-19 disease. One of the major concerns in using machine learning (ML) approaches for automatic processing of CT scan images in clinical setting is that these methods are trained on limited and biased subsets of publicly available COVID-19 data. This has raised concerns regarding the generalizability of these models on external datasets, not seen by the model during training. To address some of these issues, in this work CT scan images from confirmed COVID-19 data obtained from one of the largest public repositories, COVIDx CT 2A were used for training and internal validation of machine learning models. For the external validation we generated Indian-COVID-19 CT dataset, an open-source repository containing 3D CT volumes and 12096 chest CT images from 288 COVID-19 patients from India. Comparative performance evaluation of four state-of-the-art machine learning models, viz., a lightweight convolutional neural network (CNN), and three other CNN based deep learning (DL) models such as VGG-16, ResNet-50 and Inception-v3 in classifying CT images into three classes, viz., normal, non-covid pneumonia, and COVID-19 is carried out on these two datasets. Our analysis showed that the performance of all the models is comparable on the hold-out COVIDx CT 2A test set with 90% - 99% accuracies (96% for CNN), while on the external Indian-COVID-19 CT dataset a drop in the performance is observed for all the models (8% - 19%). The traditional machine learning model, CNN performed the best on the external dataset (accuracy 88%) in comparison to the deep learning models, indicating that a lightweight CNN is better generalizable on unseen data. The data and code are made available at https://github.com/aleesuss/c19.

# 1 Introduction

During the COVID-19 pandemic, we have seen healthcare systems of even the developed countries to be severely affected. And, an urgent need for quick and reliable tool for screening, detection, and monitoring of the disease was strongly felt. Though RT-PCR is the method of choice in the detection of COVID-19, several studies have shown that Computed tomography (CT) scans can be sensitive in the early stages of the disease diagnosis and/or complement RT-PCR with high sensitivity and also have been used as primary tools in places where RT-PCR tests were not available [1]–[3]. This led to numerous automated diagnostic solutions based on machine learning approaches proposed by researchers worldwide to assist the physicians in CT image analysis. Though deep learning (DL) models have been very reliable in many medical image data applications, these have not been very successful in the diagnosis of COVID-19 in clinical settings. Some major flaws in the methodology and/or underlying biases that have limited the applicability of these methods in a real clinical scenario has been assessed by Roberts et al in a recent review [4]. One of the major issues highlighted by the authors has been small and biased datasets used for developing and training the models. Further, an inherent bias in the publicly available datasets due to collating data from different sources, authenticity of the contributors, lack of clarity in labels as COVID-19 is not easy to diagnose, use of subsets of the original datasets without specifying the selection criteria, etc., has been pointed out as important factors in the non-reproducibility of results. Also, differences in demographics of COVID-19 cohort and the control groups, e.g., use of pediatric patients as control groups, using only internal hold-out sets for testing, developing and testing not carried out hand-in-hand with the radiologists have raised questions about the applicability of these models to the clinical setting. To address this problem, large CT image datasets covering different patient demographics are in high demand and efforts of the group led by COVID-Net CT team is commendable [5]. For deploying a model in clinical setting two important requirements are whether it is lightweight and is it reliable and generalizable. In this work, we attempt to address these objectives while taking into consideration the issues raised by Roberts et al [4]. First, we have chosen a sufficiently large and reliable dataset, COVIDx CT 2A for training. This considers only confirmed COVID-19 patient samples, from across Asian and European countries collected using a variety of CT equipment types, protocols and levels of validation. Median age of the patients in this dataset is 51 as reported in [5]. The partition of the data into training, validation and test sets were done at patient level and hence a patient is considered only in single partition. Scaling of the images were done to $224x224x3$ and this was shown to give good performance for chest radiograph classification [6]. Hence, image resolution of $224x224x3$ was used in our study. Next, for testing the generalizability of machine learning models we have generated a curated dataset of Indian patients, Indian COVID-CT dataset. Finally, to identify the simplest model with reliable sensitivity and accuracy, we considered four state-of-the-art machine models with different architectures, from a lightweight CNN with very few parameters (~2M) to very deep learning models with more and more layers and complex architecture, ranging from VGG-16 [5] to ResNet-50 [3] and Inception-v3 [4] with millions of parameters (23M – 138M). For performance evaluation, first training and validation of all the models is carried out on COVIDx CT dataset [5] and testing is carried out on two datasets, viz., the hold-out test set of COVIDx CT, and the external cohort, Indian-COVID-19 CT. To summarize, the two major contributions of this work are: (i) provide COVID-19 CT scan image dataset of Indian patients and show the usefulness of such external datasets in developing reliable AI/ML-based models for the diagnosis of COVID-19, and (ii) show that a light-weight CNN model is a better generalizable and reliable model compared to very deep models. This would provide an advantage of easy deployment on small portable devices in clinical settings.

## 2 Related Work

It has been observed that distinct patterns are observed in chest CT scans images of patients infected with COVID-19 compared to other bacterial/viral infections. Further, correlation in the patterns in CT images with severity of the disease has aided in monitoring the disease progression in COVID-19 patients. Many recent studies have shown that deep learning methods are able to capture these distinguishing patterns in CT images with comparable or better efficiency compared to expert radiologists [7]. Large open-source repositories of COVID-19 CT scan images from heterogeneous groups of patients are in high demand for the development of AI-based data driven solutions. One of the largest publicly available datasets, COVIDx CT [5], has 2D CT scan images of normal, COVID-19 and pneumonia patients collated from 17 countries (194,922 images/ 3745 patients). Numerous population-specific datasets have been proposed, a dataset of 2D images of CT scans from 282 normal and 95 COVID-19 patients from Iran [8], and other two repositories of only images from COVID-19 patients - 3D CT scans of 81 Covid positive patients from Italy [9] and COVID19-CT-dataset [10] with 1013 Covid positive patients from Iran.

COVIDNet-CT model is one of the earliest methods proposed for classifying CT scans into Normal, non-Covid Pneumonia and COVID-19 on COVIDx CT test set with an accuracy of 99.1% [5]. It uses a machine-driven design exploration strategy for building the model with ResNet architecture pre-trained on ImageNet [11]. Another study for distinguishing COVID-19 from viral pneumonia uses a pre-trained InceptionNet architecture to convert image features into a one-dimensional vector which is fed as input to a two-layered fully connected network [12]. Performance of the binary classifier was evaluated on an external validation dataset and an accuracy of 79.3%, specificity of 0.83, and sensitivity of 0.67 was reported. Performance of 10 different CNN architectures for binary classification of COVID-19 and non-COVID-19 CTs was carried out on CT images annotated by radiologists and patches of infected areas were considered for analysis in [7]. ResNet-101 resulted in a sensitivity of 100% on hold-out validation set in this study. Numerous studies have developed models for segmenting the CTs into lung field and lesions with abnormalities and using the annotated 'regions' for classification [13], [14]. These methods claim to provide better generalizability and interoperability during clinical implementations compared to original scans that are noisy and exhibit variations across different scanning devices. Studies integrating image data with other clinical data such as age, sex, exposure history, symptoms, and laboratory tests to detect COVID-19 have also been developed [14], [15]. The major limitation of some of these studies is that these are carried out on hold-out test sets and majority of them were on small datasets as data was not available in the early stages of pandemic. In some recent studies external datasets, not seen by the model during the training, have been used to evaluate the generalizability of deep learning models [16], [17], [18]. A general observation in these studies is that very deep layered models generally perform poorly on external datasets, probably due to overfitting.

## 3 Dataset Construction

A total of 288 COVID-19 patient data collected during the period April - September 2020 was obtained from Gandhi Hospital, Hyderabad, India. The manufacturer's details of the CT scanner used for image acquisition are given in **Table 1**. For analysis 42 slices that contained broad and clear lung window without any other interfering organs were chosen (in the range ~40 to 300 of dicom series) to reduce the dataset size. Each CT volume was then converted to png format. The images are plain CT scans captured with no contrast and slice thickness of the images are 0.6, 1.5, and 5 mm. Age of patients is in the range of 17 - 79 years (mean age ~ 45yrs). The 3D volumes of the data in dicom format along with 2D images in png format and instructions on how to access it is available at https://github.com/aleesuss/c19. To the best of our knowledge, there is no publicly available CT image dataset from the Indian population and the proposed Indian-COVID-19-CT dataset is one of its kind from India.

**Table 1.** Details of the CT scanner machine used to acquire Indian-COVID-19 CT dataset

| Key | Value |
|---|---|
| Manufacturer | SIEMENS |
| Modality | CT |
| Manufacturer’s Model Name | Emotion 16 |
| Device Serial Number | 39306 |
| Software Version(s) | Syngo CT 2014A |

For training the publicly available benchmark dataset, COVIDx CT, is used. It includes CT images of normal, pneumonia, and COVID-19 classes, collated from multiple data sources worldwide. It consists of 194922 CT images from 3745 patients of which 94,548 images are from 2299 COVID-19 patients. It has been split into 60-20-20 ratio for training, validation and testing the models, and summarized in **Table 2**. The Indian-COVID-19 CT dataset, is used as an external test set of Covid class for evaluating the generalizability of the various DL architectures.

**Table 2.** Number of images (patients) in the three classes in COVIDx CT dataset used for training, validation and testing are given. Details of external test set, Indian-COVID-CT, is also given. *Both test sets in hold out and external datasets used same Normal and Pneumonia images.

| Type | Normal | Pneumonia | Covid | Total |
|---|---|---|---|---|
| **COVIDx CT dataset:** | | | | |
| Train | 35996 (321) | 25496 (558) | 82286 (1958) | 143778 (2837) |
| Validation | 11842 (126) | 7400 (190) | 6244 (166) | 25486 (482) |
| Test | 12245 (126) | 7395 (125) | 6018 (175) | 25658 (426) |
| **Indian-COVID-CT dataset:** | | | | |
| Test | 12245 (126)* | 7395 (125)* | 12,096 (288) | 12,096 (288) |

# 4 Model Architecture

Convolutional Neural Network models are very popularly used for image classification tasks and many CNN based DL models were proposed to improve on the performance of earlier models. VGG-16 was proposed in 2014, followed by ResNet50 and Inception-v3 in 2016, the details of each are given below. **Fig. 1** gives an overview of CNN and the CNN based DL models used in this study.

## 4.1 CNN

The architecture of the CCN model used in this study to classify chest CTs into three classes, viz., normal, non-covid pneumonia and COVID-19, is given in **Fig. 1** It consists of 6 convolutional blocks with each block comprising two convolution layers, with the convolution operation represented by the expression:

$$G[m,n] = (f * h)[m,n] = \Sigma_j \Sigma_k h[j,k] f[m-j, n-k] \quad (1)$$

where f represents the input image, h, the kernel function, and $m \times n$ is the size of the convolution matrix. The first block has 16 filters followed by 32, 64, 128, 256, and 512 filters in successive blocks. All kernels are of size $3 \times 3$ and a zero padding is used to make the input and output width and height dimensions the same. A ‘maxpool’ layer is added after first convolution block and a ‘batch normalization’ followed by ‘maxpool’ layer added for the remaining five convolutional blocks. ‘Maxpool’ layer is used for down sampling the resultant matrices from the convolution operation by selecting the most dominant pixel out of a set of neighboring pixels defined by the size $m \times n$. A dropout layer is added after fourth, fifth and sixth convolutional blocks to avoid overfitting. The convolutional blocks are followed by dense layers with 512, 128, 64 and 3 nodes in each layer. Dropout layers are also used after each dense layer. The output layer has a ‘softmax’ activation function, $f(z) = 1/(1 + e^{-z})$, and previous layers of convolution and dense layers used ‘Relu’ function, $f(z) = \max(0, z)$, where $z$ is the input signal. The 'softmax' function in the output layer computes the probability distribution of the output classes as $o^k = e^{x^k} / \sum_1^n e^{x^n}$, where $x$ is input vector and $o$ is output vector and sum of all outputs is equal to 1. The loss function used was *categorical cross entropy*. The input image dimensions are $224 \times 224 \times 3$ and batch size of 8 was chosen based on available computational resources. To find optimal number of convolution blocks, two experiments were performed using 5 & 6-layered CNN architectures, repeated 3 times for each case and averaged results are given in **Table 3**. Though training accuracy increased from 97.8 to 99, validation accuracy dipped from 97.25 to 95.67, and test accuracy, 96.3, was comparable. This indicates the model may have started overfitting and we did not further increase the number of blocks. Since adding a dense layer increased the number of parameters from 2.9M to 3.7M for the 6-layered architecture and the training accuracy was quite good (~99), no further changes were made in the number of dense layers. For choosing optimal number of epochs, 4 experiments were performed with 10, 18, 20, and 30 epochs for the 6-layered CNN model and results are given in **Table 4**. To select best validated model, after each epoch during the training phase, performance of the model was evaluated on validation set and weights of the model were saved if accuracy of the model improved. After the training phase, weights of the model that gave best accuracy across all the epochs were considered in the testing phase. Number of nodes in first dense layer were chosen based on number of filters in previous Conv block and halved in subsequent dense layers. Number of dropout layers was chosen empirically.

All the DL models used in this study were pre-made and taken from 'Keras Applications' library [22].

### 4.2 VGG-16

The VGG-16 architecture had five blocks of convolutional layers with kernels of 3x3 size followed by three fully-connected layers. A max pool layer with 2 × 2 kernels is used after each block. The fully connected layers of VGG-16 were removed and a customized fully connected block with four layers each having 512, 128, 64 and 3 nodes in each layer was added. Relu activation function was used for all fully connected layers except for the last layer where 'softmax' function was used to classify images into 3 classes.

### 4.3 ResNet50

The ResNet50 architecture had 16 convolution blocks with 3 convolution layers each followed by one layer of fully connected nodes. The shortcut connections with the 'Residual blocks' are used for improving efficiency by overcoming the 'vanishing gradient' problem in deep neural networks. The fully connected layer in ResNet50 was also replaced with a fully connected layer with 3 nodes with 'softmax' activation function.

### 4.4 Inception-v3

The Inception-v3 architecture was a 42 layered deep network with 3 inception modules separated by grid size reduction modules for feature map downsizing. Each inception module consists of multiple convolution layers. In this model also the same fully connected module as in VGG-16 was used. VGG-16, ResNet50 and Inception-v3 were pre-trained using ImageNet before training using COVIDx CT dataset.

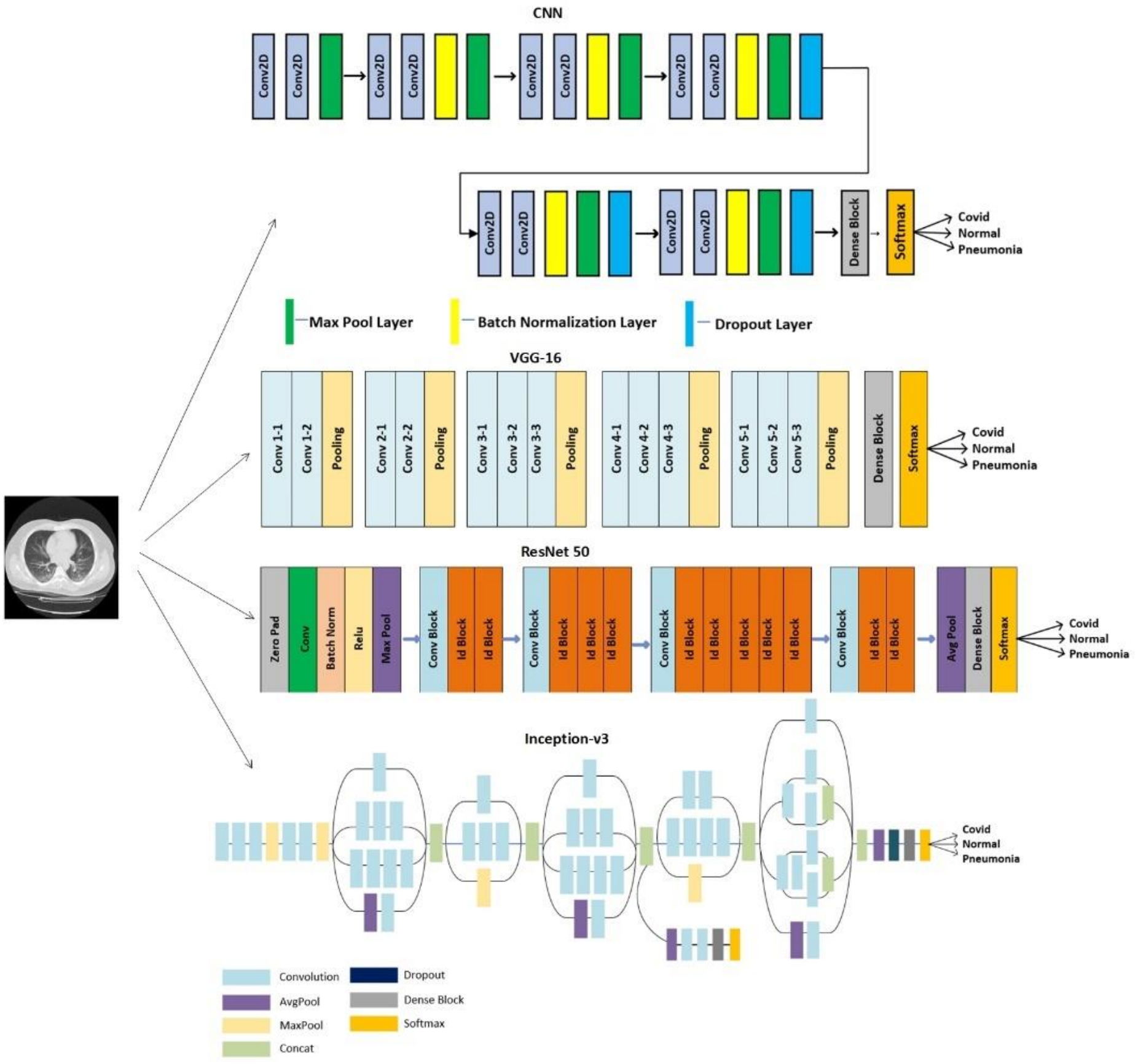

**Fig. 1.** Representative architectures of the models used in this paper.

**Table 3.** Average accuracies from three experimental runs with 5 and 6 layered CNN models on COVIDx CT dataset for 10 epochs.

| No. of Layers | Training | Validation | Testing |
|---|---|---|---|
| 5 | 97.8 | 97.25 | 96.1 |
| 6 | 99.0 | 95.67 | 96.3 |

**Table 4.** Accuracies of the experiments repeated with COVIDx–CT dataset for 10, 18, 20 and 30 epochs for 6 layered CNN model.

| No. of Epochs | Training | Validation | Testing |
|---|---|---|---|
| 10 | 99.0 | 95.0 | 95.0 |
| 18 | 99.44 | 96.11 | 95.97 |
| 20 | 99.82 | 96.11 | 96.68 |
| 30 | 99.73 | 96.20 | 96.78 |

# 5 Implementation

All the models were trained on 4 GeForce GTX 1080 Ti GPUs. The time taken for training the CNN model was ~ 62 hours for 30 epochs, the DL models took from 17 hours to 4 days for completing 3 epochs. The optimizer used was Adam with an initial learning rate set to 5e-5 for CNN and 0.0001 for other models. It was set to reduce by 0.3 if no improvement in validation loss was observed for 2 epochs and decay rate of first and second moments were set to the default values of 0.9 and 0.999, respectively. The cyclic learning rate policy proposed in [23] was followed to decide the learning rate for CNN. First the learning rate was set randomly to the default value of 0.001, then reduced manually in each trial to identify optimal learning rate. For default value of 0.001, accuracy oscillated between 0.46 and 0.28. Heuristically reducing learning rate to 5e-3 also did not result in any significant improvement. Further reducing to 5e-5, model performance improved to 0.97, but dropped on further reducing learning rate to 5e-6 in next trial. So, learning rate was fixed at 5e-5. Our analysis revealed that the lower and upper bounds of the optimal learning rate for this system lies between 5e-3 and 5e-6. The versions of Python, Tensorflow and other libraries used in the implementation are given in github link, https://github.com/aleesuss/c19.

# 6 Results

First, we present performance evaluation of the 5 models on the hold-out test of COVIDx CT dataset. The accuracy and loss curves for the CNN model is given in **Fig. 2** and the evaluation metrics, viz., Precision, Recall, F1-score and accuracy of all the models are summarized in **Table 5**. The CNN model was trained for 30 epochs on COVIDx CT dataset and training and validation accuracies converged to 0.99 and 0.96 respectively. It is observed that both the accuracy and loss curves plateau after 7 epochs for the CNN model, indicating the model has stabilized without over-fitting. It is observed that the three DL models achieved high accuracy by 3 epochs. From **Table 5** it is observed that ResNet-50 outperformed with precision and recall of (0.98, 0.98) for the Covid class, followed by the CNN model with comparable values (0.96, 0.94), VGG-16 (0.89, 0.89), Inception-v3 (0.82, 0.84). It is worth noting that the CNN model used much fewer training parameters, ~2M, compared to the DL models: ResNet-50 - 23M, Inception-v3 - 24M, VGG-16 - 138M.

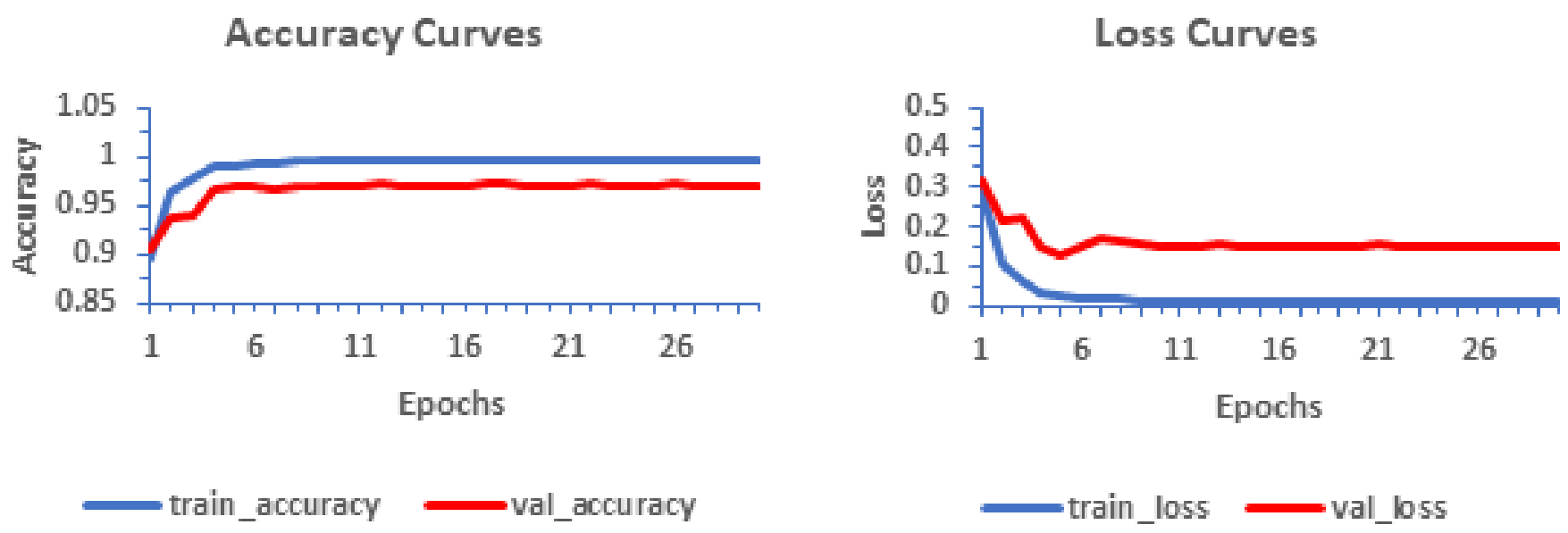

**Fig. 2.** Accuracy and Loss Curves for the CNN model is shown.

**Table 5.** Comparative performance of the four machine learning models on COVIDx CT test data.

| Model | Class | Precision | Recall | F1-score | Accuracy |
|---|---|---|---|---|---|
| **CNN** | Covid | 0.96 | 0.94 | 0.95 | 96% |
| | Normal | 0.96 | 0.98 | 0.97 | |
| | Pneumonia | 0.99 | 0.97 | 0.98 | |
| **VGG-16** | Covid | 0.89 | 0.89 | 0.89 | 94% |
| | Normal | 0.96 | 0.96 | 0.96 | |
| | Pneumonia | 0.94 | 0.94 | 0.94 | |
| **ResNet50** | Covid | 0.98 | 0.98 | 0.98 | 99% |
| | Normal | 0.99 | 0.99 | 0.99 | |
| | Pneumonia | 0.99 | 1.00 | 0.99 | |
| **Inception-v3** | Covid | 0.82 | 0.84 | 0.83 | 90% |
| | Normal | 0.96 | 0.95 | 0.95 | |
| | Pneumonia | 0.89 | 0.88 | 0.88 | |

To assess the generalizability of the models considered, we next evaluated their performance on the external cohort, Indian-COVID-19 CT dataset and the results are summarized in **Fig. 3**. For this experiment, the normal and pneumonia CT images were taken from COVIDx CT test set (see **Table 1**). A consistent drop in the performance of all the four models is observed for the COVID class. This is not surprising as this is an external cohort, not seen by the model during training. The CNN model outperformed with an accuracy of 0.88, while that of ResNet-50 dropped to 0.81. From the confusion matrix for CNN model given in **Table 6**, we observe that though majority of COVID-19 cases were correctly identified by the model resulting in high precision (0.96), while a significant number of cases were predicted as Normal, resulting in low recall value (0.75). The recall values for the COVID class for the three DL models are much lower as seen in **Fig. 3**. The 95% confidence intervals of the precision and recall values were calculated and were found to be within [1.5, 1.3] for all the classes. These results show that the lightweight CNN model with fewer parameters generalized much better compared to the very deep models.

## 7 Discussion

One of the major contributions of this work is the construction of curated dataset of CT scan images of COVID-19 from India. Population-specific datasets are highly desired for assessing the generalizability of deep learning (DL) models. These can also be used in automated generation of datasets using Generative Adversarial Networks (GANs) to address the problem of limited data for training DL models [24]. Other applications of the Indian-COVID-19 CT dataset include training ML algorithms for the detection of lung abnormalities in general, and in developing applications for segmentation of lungs and infected regions at the slice level. Slice level classification models based on the presence or absence of “markers for infection” have been applied for detecting various lung diseases [25]. Given the importance of this dataset, care has been taken in its construction following the recommendations proposed in [4]. For example, all images in this dataset are from the same scanner and confirmed through reliable sources to include only COVID-19 positive patients. The current limitation of the dataset is that it contains CT images from only COVID-19 patients and the data size is small.

The second objective of this work is to propose a lightweight model that can be easily deployed on portable machines in a clinical setting. For this, performance of a lightweight CNN model with a simple architecture is compared with three CNN-based deep learning models. The performance of the four models is evaluated on both, the hold-out data from COVIDx CT (used for training and validation) and on external Indian COVID-CT dataset (not seen by the models during training phase) to show the generalizability of these models.

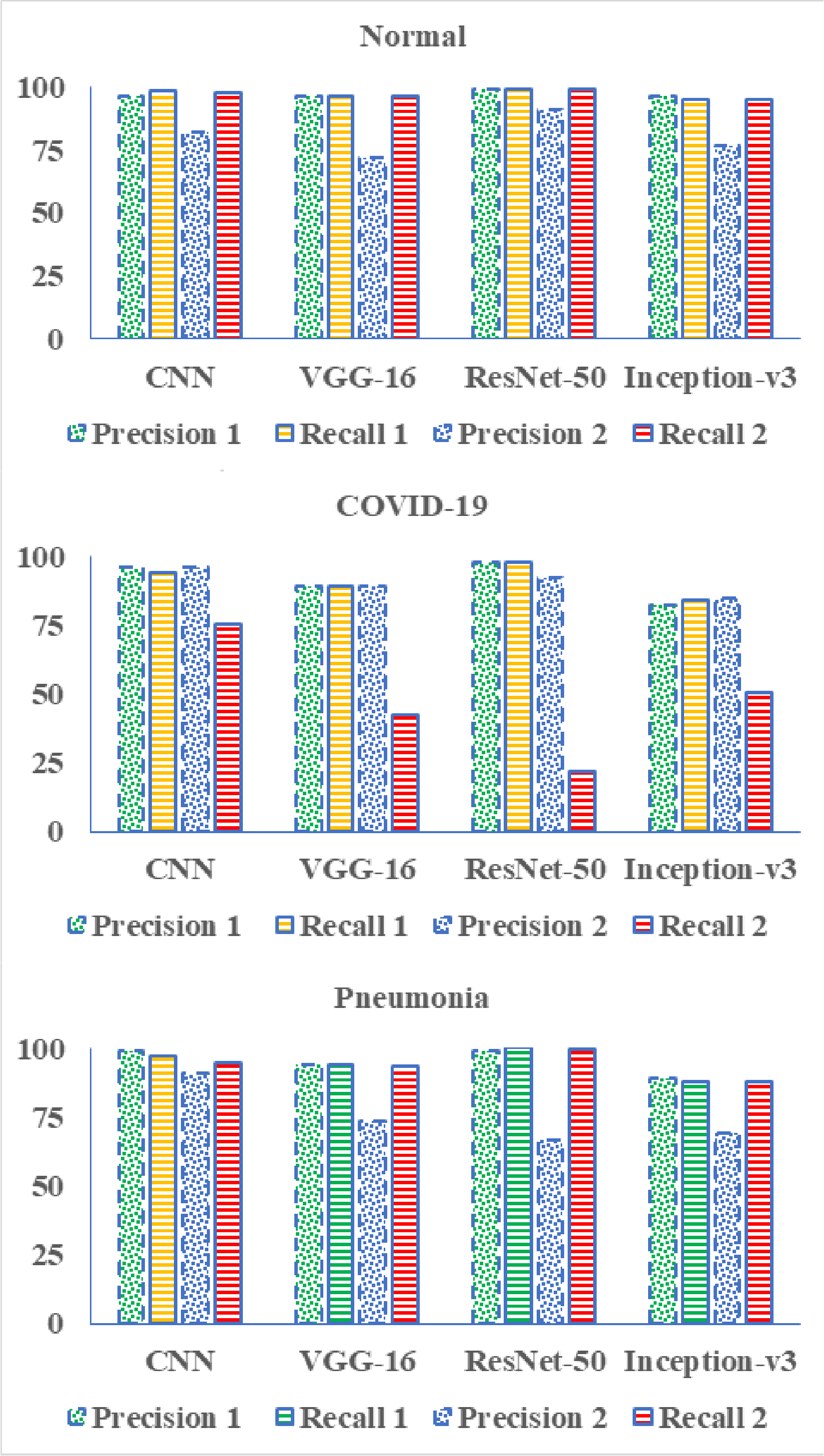

**Fig. 3.** Precision and Recall values of the four ML models on the holdout set of COVIDx CT dataset (Precision 1 and Recall 1) and external Indian-COVID-19 CT (Precision 2 and Recall 2) for the COVID class is shown. Normal and pneumonia images were taken from COVIDx CT.

**Table 6.** Confusion Matrix on testing CNN model's performance on Indian-COVID-19 CT data.

| | **Covid-19** | **Normal** | **Pneumonia** |
|---|---|---|---|
| **Covid-19** | 9092 | 2334 | 670 |
| **Normal** | 264 | 11954 | 27 |
| **Pneumonia** | 88 | 302 | 7005 |

The reason for proposing a lightweight CNN model for the detection of COVID-19 is that the imaging patterns found in CT images of COVID-19 patients can basically be associated with three patterns of pneumonia findings, viz., peripheral, multifocal and diffuse patterns [26]. Salehi et al. reports the frequencies of the different CT abnormalities seen in COVID-19 patients as follows: ground glass opacification (GGO) (88.0%), bilateral involvement (87.5%), peripheral distribution (76.0%), and multilobar (more than one lobe) involvement (78.8%) and consolidation (31.8%) [27]. From these findings it could be deduced that there are only a few characteristic features of COVID-19 such as GGOs, consolidations and in severe cases, crazy paving patterns, typically distributed peripherally in multiple lobes of the lungs. Also, COVID-19 pneumonia presentation is very different from pneumonia of bacterial and other origins, with an absence of centrilobular nodules and no mucoid impactions in the absence of superinfection [28]. Not much variation in the CT images of COVID-19 pneumonia patients which are easily captured by a simple CNN architecture, and a very deep learning model may not be necessary as there may be a possibility of overfitting with DL models. As seen in **Table 5** and **Fig. 3**, poorer performance of all the three DL models on the external dataset confirms that a CNN model with fewer parameters may suffice for this task. Further, its performance on unseen data suggests it to be a good generalizable model and reliable in a clinical setting. We observe that though ResNet-50 exhibited best performance on COVIDx CT test set, followed by our CNN model, its performance reduced with a very low recall value of 0.22, though precision was good at 0.92. This shows that learning using very deep models on limited datasets can lead to overfitting and make it less generalizable. This is in agreement with similar observations made in other studies [16], [18]. In [16], Nguyen et.al examined the generalizability of nine different DL models using data from external dataset other than the one used for training and reported fall of performances of models close to an AUC of 0.5 when tested on external datasets. Wynants et.al [18] reports high bias associated with COVID-19 prediction models using prediction model risk of bias assessment tool (PROBAST). This does not imply that the models were not trained properly, but rather due to other factors such as the external dataset may have different patient demographics, data collection protocols across different labs, etc. leading to variation in the training and external test sets. Thus, building a large collection of CT image datasets from different population groups would be of great help in model development along with continuous learning.

Further, from the confusion matrix we observe that for all the models, many COVID-19 images were classified as normal and resulted in lower recall values for the COVID class in **Fig. 3**. One possible reason for this could be variation in severity of the disease across patients, and so some of the chest CT images may not exhibit any of the characteristic features of COVID-19 such as ground-glass opacities, consolidation and crazy lines. Many patients do not elicit the pulmonary inflammatory response needed to produce the chest CT findings of lung injury [29]. Since negative CT result does not rule out COVID-19 infection, this has led to the concern of usage of CT scans for the detection of COVID-19 by most radiological societies to avoid unnecessary exposure to radiation. However, since CT scans are very sensitive in distinguishing between different types of pneumonia (bacterial, viral, COVID) and given that CT scans have fewer false negatives (~9%) [30] and give immediate results compared to RT-PCR, these have proven very useful during the pandemic to assess the severity of the disease in patients. Thus, under such scenario, generalizable automated diagnosis tools for detecting the cause of pneumonia using machine learning models would be an asset for doctors in screening the patients.